\documentclass{elsart}
\usepackage{epsfig}
\begin{document}

\begin{frontmatter}

\title{The kaon electromagnetic form factor}

\author{J. Lowe}
\address{Physics and Astronomy Department, University of New
Mexico, Albuquerque, \\NM 87131, USA}

\author{M.D. Scadron}
\address{Physics Department, University of Arizona, Tucson, 
AZ 85721, USA}

\begin{abstract}
We use recent data on $K^+\rightarrow\pi^+e^+e^-$, together
with known values for the pion form factor, to derive experimental
values for the kaon electromagnetic form factor for 
$0<q^2<0.125~{\rm (GeV/c)}^2$. The results are then compared with
predictions of the Vector-Meson-Dominance model, which gives a
good fit to the experimental results.
\end{abstract}
\end{frontmatter}

\section{Introduction}
  
The pion electromagnetic form factor, $F_{\pi}(q^2)$, is well studied
experimentally. Many measurements, for both positive and negative 
$q^2$, have been reported in the literature \cite{radii}. In addition
to direct measurements of $F_{\pi}(q^2)$, further
information comes from the pion charge radius, which is related to
the slope of the form factor at $q^2=0$. By contrast, much less
is known about the kaon form factor, $F_K(q^2)$. There are some
measurements \cite{radii} for negative $q^2$. For positive $q^2$,
the only measurements \cite{new} are for $q^2>1~{\rm (GeV/c)}^2$,
leaving the region $0<q^2<1~{\rm (GeV/c)}^2$ unexplored.
 
Information on $F_K$ can be deduced from experimental
data on the decay $K^+\rightarrow\pi^+e^+e^-$, such as that 
provided by the recent high-statistics Brookhaven experiment 
E865 \cite{865}. The amplitude for this decay was measured for $q^2$
up to 0.125 (GeV/c)$^2$, the maximum allowed by the kinematics of this
kaon decay. This amplitude does not give $F_K$ directly, but
rather the difference $F_K(q^2)-F_{\pi}(q^2)$. Since $F_{\pi}(q^2)$ is 
relatively well known, then, this decay is a source of information on 
$F_K$.
 
In the present paper, we extract $\mid F_K(q^2)\mid^2$ from the
E865 results for $K^+\rightarrow\pi^+e^+e^-$ and compare the 
results with the predictions of the Vector-Meson-Dominance (VMD)
model \cite{vmd}.
 
\section{The decay $K^+\rightarrow\pi^+e^+e^-$ and the kaon
form factor}
 
The decay $K^+\rightarrow\pi^+e^+e^-$ has been studied theoretically
for many years. Already in 1985 it became clear \cite{es} that
the process is dominated by the ``long-distance" (LD) terms, in
which a virtual photon is radiated by either the pion or the kaon.
However, it was not until the detailed data of experiment E865
\cite{865} became available that a convincing description of
both the scale and the $q^2$ dependence of the amplitude was found
\cite{bels}.
 
\begin{figure}[t]
\hspace{-1.5cm}
\epsfig{file=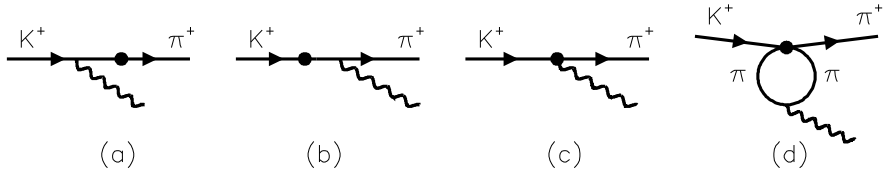,width=15.5cm}
\caption{\label{fig.1} Graphs for $K^+\rightarrow\pi^+e^+e^-$. 
(a) and (b) are
long-distance graphs, (c) is a short-distance graph and (d) is the
pion loop term. In each graph, the blob denotes the weak 
(strangeness-changing) vertex. The (off-shell) photon converts
to $e^+e^-$.}
\end{figure}

Burkhardt et al. \cite{bels} considered four contributions
to the amplitude, depicted in Fig. \ref{fig.1}. The LD terms are
those in Fig. \ref{fig.1}(a) and (b). These two graphs are related
to the pion and kaon form factors as shown below. Fig. 
\ref{fig.1}(c) represents all short-distance (SD) terms. These were
already known in ref. \cite{es} to be small, and subsequent work 
\cite{ddg} has shown them to be still smaller than previously
believed.
Therefore here, as in ref. \cite{bels}, we neglect the SD
contribution from Fig. \ref{fig.1}(c). Fig. \ref{fig.1}(d) is a 
``pion loop" term, first discussed by Ecker et al.
\cite{piloop}. Its contribution is small, but it gives
a characteristic shape to the $q^2$ dependence of the amplitude.
The data of ref. \cite{865} and the analysis of ref. \cite{bels}
each show the ``kink" at $q^2\sim (2m_{\pi})^2$ resulting from
this pion loop term, providing convincing evidence for the presence
of the term. As in ref. \cite{bels}, we take this term directly
from \cite{piloop}. 
 
The pion and kaon form factors enter {\it via} the graphs of
Fig. \ref{fig.1}(a) and (b), which give the LD amplitude 
\cite{bels} 
\begin{eqnarray}
\mid A_{LD}(q^2)\mid=e^2\bigg|
\frac{\langle\pi^+\mid H_W\mid 
K^+\rangle}{m^2_{K^+}-m^2_{\pi^+}}\bigg|~~\bigg|
\frac{F_{\pi^+}(q^2)-F_{K^+}(q^2)}{q^2}\bigg|.
\label{eq.1}
\end{eqnarray}
 
\noindent In the numerical calculations below, we take the value
of the weak matrix element $\langle\pi^+\mid H_W\mid K^+\rangle$
from ref. \cite{ls}, following from the average over {\it eleven}
different measures of weak-decay matrix elements:
\begin{eqnarray}
\mid\langle\pi^+\mid H_W\mid K^+\rangle\mid =(3.59\pm 0.05)
\times 10^{-8}~{\rm GeV}^2.
\label{eq.2}
\end{eqnarray}
 
\noindent Adding the pion loop amplitude, Fig. \ref{fig.1}(d),
from Ecker et al. \cite{piloop}, we obtain

\begin{eqnarray}
A(q^2)&=&A_{LD}(q^2) + A_{\pi loop}(q^2)
\nonumber \\
 &=&e^2\bigg|\frac{\langle\pi^+\mid H_W\mid K^+\rangle}
{m^2_{K^+}-m^2_{\pi^+}}\bigg|~~\bigg|
\frac{F_{\pi^+}(q^2)-F_{K^+}(q^2)}{q^2}\bigg|+
A_{\pi loop}(q^2)
\label{eq.3}
\end{eqnarray}
 
\noindent from which

\begin{eqnarray}
\mid F_K-F_{\pi}\mid = \frac{q^2(m^2_{K^+}-m^2_{\pi^+})}
{e^2\mid\langle\pi^+\mid H_W\mid K^+\rangle\mid}
\bigg[A(q^2)-A_{\pi loop}(q^2) \bigg].
\label{eq.4}
\end{eqnarray}

To apply Eq. (\ref{eq.4}), we need experimental values for
$A(q^2)$ and $F_{\pi}(q^2)$. For the former, we use data from
Brookhaven E865 \cite{865}. Their amplitude $f(q^2)$ is related
to our $A(q^2)$ by
 
\begin{eqnarray}
\mid A(q^2)\mid=f(q^2)\frac{G_F\alpha}{4\pi}
\label{eq.5}
\end{eqnarray}
 
\noindent where $G_F$ is the Fermi constant and $\alpha$ is the fine
structure constant. 

\begin{figure}[h]
\hspace{4.5cm}
\epsfig{file=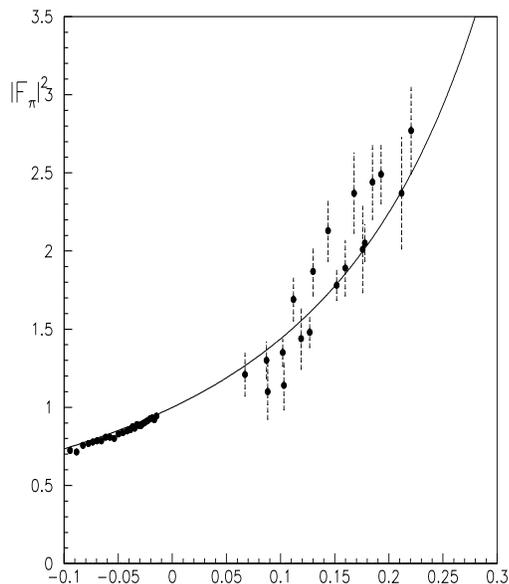,height=9.1cm,width=7.2cm}
\caption{\label{fig.2} Pion electromagnetic form factors squared.
The points are the experimental data and the solid line is the
VMD prediction.}
\end{figure}

The experimental values of $F_{\pi}(q^2)$ from ref. \cite{radii}
are in general not measured at precisely the required values of
$q^2$. However, the data, which are plotted in Fig. \ref{fig.2},
are well described for the region of $q^2$ of interest by the VMD
model using a rho-meson pole:

\begin{eqnarray}
F_{\pi}^{VMD}(q^2)= \frac{m^2_{\rho}}{m^2_{\rho}-q^2}
\label{eq.6}
\end{eqnarray}

\noindent  where \cite{PDG} $m_{\rho}=775.8~{\rm MeV}$. This is
shown by the solid line in Fig. \ref{fig.2}. A quantitative test
of the agreement between the VMD and the experimental data is
provided by the value of $\chi^2$, defined by
 
$$\chi^2=\Sigma\bigg[\frac{\mid F^{exp}_{\pi}\mid^2-
\mid F^{VMD}_{\pi}\mid^2}{\sigma}\bigg]^2$$

\noindent where the sum runs over the 14 points for $q^2>0$,
$\mid F^{exp}_{\pi}\mid^2$ and $\mid F^{VMD}_{\pi}\mid^2$ are 
respectively the
experimental values and the VMD predictions, and $\sigma$ is the
experimental error on $\mid F^{exp}_{\pi}\mid^2$. There are no
variable parameters to search on, so the number of degrees of
freedom is 14. The result is $\chi^2/{\rm degree~of~freedom}=0.89$,
which shows that the VMD curve reproduces the data satisfactorily
in the relevant region of $q^2$. Therefore we use Eq. 
(\ref{eq.6}) to calculate the required values of $F_{\pi}$. 
We emphasise that the values we calculate this way are essentially
experimental; the VMD prediction is used only as an
interpolating function between the measured experimental points.

The relative sign of $A_{LD}$ and $A_{\pi loop}$ was already
established in refs. \cite{865} and \cite{bels}. To derive
$F_K$ from Eq. (\ref{eq.4}), we need also to determine the
sign of $F_K-F_{\pi}$. To do so, we observe that in the VMD model,
and in other models such as all those discussed by Scadron
et al. \cite{klee}, as well as in all the available
data for $q^2<0$, $F_K$ differs less from unity than does 
$F_{\pi}$, i.e. $$\mid F_K-1\mid ~<~\mid F_{\pi}-1\mid.$$
We assume that this inequality holds also for 
$0<q^2<0.125~{\rm (GeV/c)}^2$. This defines the required sign,
giving
 
\begin{eqnarray}
F_K=F_{\pi}-\frac{q^2(m^2_{K^+}-m^2_{\pi^+})}
{e^2\mid\langle\pi^+\mid H_W\mid K^+\rangle\mid}
\bigg[A(q^2)-A_{\pi loop}(q^2)\bigg].
\label{eq.7}
\end{eqnarray}

\begin{figure}[h]
\hspace{4.5cm}
\epsfig{file=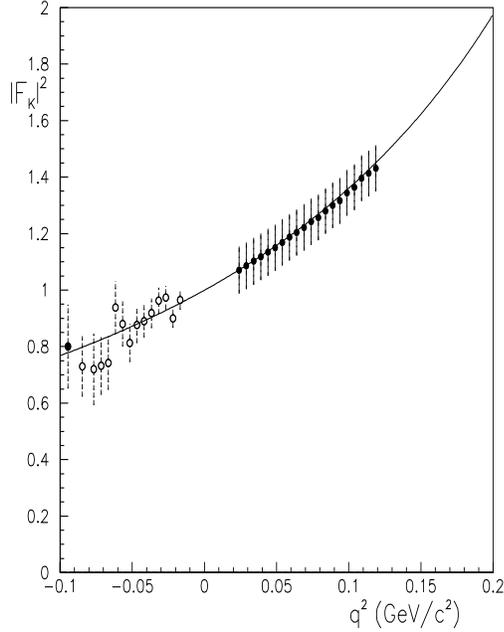,height=9.4cm,width=7.2cm}
\caption{\label{fig.3} Kaon electromagnetic form factors squared.
The solid points are the experimental data from the present
analysis and the circles show the previously existing data. The 
solid line is the VMD prediction.}
\end{figure}

The extracted values of $\mid F_K\mid^2$ are listed in Tab.
\ref{tab.1} and plotted in Fig. \ref{fig.3} which also shows
previous data, from ref. \cite{radii}, for $q^2<0$. The errors
in our values of $F_K$
arise from experimental errors in the amplitude $A(q^2)$ for 
$K^+\rightarrow\pi^+e^+e^-$ and the error in the weak matrix
element $\langle\pi^+\mid H_W\mid K^+\rangle$, but dominantly
from the errors on the experimental values of $F_{\pi}$. The VMD
model, Eq. (\ref{eq.6}), gives a reasonably unambiguous value
for $F_{\pi}$. However, our input values for $F_{\pi}(q^2)$ 
are basically experimental, and the use of VMD as an interpolating
function is only justified to the extent that
it agrees with the experimental points in Fig. \ref{fig.2}. The
weighted RMS deviation of the positive-$q^2$ experimental points
in Fig. \ref{fig.2} from the VMD prediction is 0.0837, and we
take this as the error in $\mid F_{\pi}\mid^2$.

\begin{center}
\begin{table}
\caption{\label{tab.1} Experimental values for the kaon form
factor from the present analysis.}
\vspace{0.3cm}
\center{
\begin{tabular}{cc||cc}   
\hline
$q^2$ & $\mid F_K(q^2)\mid^2$ & $q^2$ & $\mid F_K(q^2)\mid^2$ \\
${\rm (GeV/c)}^2$ & & ${\rm (GeV/c)}^2$ & \\ \hline
0.0244 & $1.071\pm 0.083$ & 0.0744 & $1.242\pm 0.082$ \\
0.0294 & $1.086\pm 0.083$ & 0.0794 & $1.257\pm 0.082$ \\
0.0344 & $1.102\pm 0.083$ & 0.0844 & $1.280\pm 0.081$ \\
0.0394 & $1.118\pm 0.083$ & 0.0894 & $1.300\pm 0.081$ \\
0.0444 & $1.134\pm 0.083$ & 0.0944 & $1.316\pm 0.081$ \\
0.0494 & $1.150\pm 0.082$ & 0.0994 & $1.344\pm 0.081$ \\
0.0544 & $1.169\pm 0.082$ & 0.1044 & $1.364\pm 0.081$ \\
0.0594 & $1.187\pm 0.082$ & 0.1094 & $1.396\pm 0.081$ \\
0.0644 & $1.204\pm 0.082$ & 0.1144 & $1.413\pm 0.081$ \\
0.0694 & $1.222\pm 0.082$ & 0.1194 & $1.431\pm 0.082$ \\ \hline
\end{tabular}
}
\end{table}
\end{center}
 
\section{Vector meson dominance}

Since the VMD picture gives a good description of the pion
form factor, as shown in Fig. \ref{fig.2}, it is of interest 
to test it also for the kaon form factor. As discussed in
sect. 2 above, the pion form factor is dominated in this model by
the $\rho$ pole, Eq. (\ref{eq.6}).
For the kaon form factor, there are contributions from
the $\rho$, $\omega$ and $\phi$ poles:
 
\begin{eqnarray}
F_K^{VMD}(q^2)=N\bigg(\frac{1}{2}
\frac{g_{\rho ee}}{m^2_{\rho^0}-q^2}
+\frac{1}{2}\frac{g_{\omega ee}}{m^2_{\omega}-q^2}+
\sqrt{\frac{1}{2}}\frac{g_{\phi ee}}{m^2_{\phi}-q^2}\bigg).
\label{eq.8}
\end{eqnarray}
 
\noindent where $g_{\rho ee}=4.97$, $g_{\omega ee}=17.06$ and 
$g_{\phi ee}=13.38$, derived from the decay widths. The masses
in Eq. (\ref{eq.8}) are taken from ref. \cite{PDG}. The
$\rho^0K^+K^-$, $\omega K^+K^-$ and $\phi K^+K^-$ SU(3) 
coefficients are $1/2$, $1/2$ and $1/\sqrt{2}$ respectively. The
requirement that $F(0)=1$ gives the normalisation coefficient
as $N=0.03682~{\rm GeV}^2$. 
 
The prediction of Eq. (\ref{eq.8}) is plotted with the data
for $\mid F_K\mid^2$ in Fig. \ref{fig.3}. As for 
$\mid F_{\pi}\mid^2$, the VMD gives excellent agreement with
data, both for the previously available data for $q^2<0$ and
for the new data derived in the present paper.

A further check on the VMD model is provided by the charge radius, 
$r$, which is related to the form factor by \cite{klee}
 
\begin{eqnarray}
r\equiv\sqrt{\langle r^2\rangle}=\hbar c
\sqrt{6\bigg[\frac{dF(q^2)}{dq^2}\bigg]_{q^2=0}}.
\label{eq.9}
\end{eqnarray}

\noindent The quantity $dF(q^2)/dq^2$ is straightforwardly
obtained from Eq. (\ref{eq.8}) giving

\begin{eqnarray}
r_K^{VMD}=\hbar c\sqrt{6N\bigg(\frac{1}{2}
\frac{g_{\rho ee}}{m^4_{\rho^0}}
+\frac{1}{2}\frac{g_{\omega ee}}{m^4_{\omega}}+
\sqrt{\frac{1}{2}}\frac{g_{\phi ee}}{m^4_{\phi}}\bigg)}
\label{eq.10}
\end{eqnarray}

\noindent with $\hbar c\approx 197.3~{\rm MeV~fm}$.
The result is $r^{VMD}_K=0.574~{\rm fm}$, which is in good
agreement with the experimental value \cite{PDG} 
$r^{exp}_K=(0.56\pm 0.03)~{\rm fm}$.
 
\section{Vector and axial-vector form factors}

Returning to meson form factors generated from 
$\pi^-,K^-\rightarrow e\nu\gamma$ decays, p. 498 of the PDG 
tables \cite{PDG} gives the vector and axial vector charged
pion form factors as $F_V(0)=0.017\pm 0.008$, 
$F_A(0)=0.0116\pm0.0016$. However, the ratio of axial vector to
vector form factors at $q^2=0$ is \cite{bram} 
$\gamma^{LsM}=1-1/3=2/3$
due to the Linear $\sigma$ Model quark plus meson loops for
$\pi\rightarrow e\nu\gamma$ decay. Then, using 
$f_{\pi}\approx 93~{\rm MeV}$, the theoretical CVC estimates
\cite{vaks} of these pion form factors are, for charged pion 
mass 139.57 MeV, $F_V(0)=m_{\pi}/8\pi^2 f_{\pi}=0.0190$,
$F_A(0)=m_{\pi}/12\pi^2 f_{\pi}=0.0127$, in reasonable agreement
with the pion form-factor data above.
 
  As for the $K\rightarrow e\nu\gamma$ decay, p. 621 of the
PDG tables \cite{PDG} gives the form factor {\it sum} as
$|F_A+F_V|=0.148\pm 0.010$, whereas refs. \cite{bram,karl} find
in the linear sigma model 0.109+0.044=0.153, compatible with 
the data above. This sum is for the kaon form factor, which is
the main subject of this paper. For the kaon charge
radius, the SU(3) analogue of the VMD SU(2) value 
$r_{\pi}=\sqrt{6}/m_{\rho}$ is then 
$r_K \approx \sqrt{6}/m_{K^*}\approx 0.54~{\rm fm}$. However the
exact VMD kaon charge radius of $0.574~{\rm fm}$ can only be
found from Eq. (\ref{eq.10}).
 
\section{Summary}
 
We have derived new values for $\mid F_K\mid^2$ for positive
$q^2$ from the experimental amplitude for the decay 
$K^+\rightarrow\pi^+e^+e^-$. Both the new values of
$\mid F_K\mid^2$, as well as the previous results for $q^2<0$,
agree well with the VMD prediction, as does the experimental
value of the kaon charge radius. This picture is consistent
with the conclusions of Ivanov et al. \cite{new} who measured
$F_K$ at $q^2>1~{\rm GeV}^2$, substantially higher than in the
present work. They found that VMD gives a good fit to their
data, at least at the lower end of their range of $q^2$.
               
\ack{ 
We acknowledge support from the US DOE.}

\end{document}